\documentclass[preprint,12pt,aps,amsmath,amssymb,nofootinbib,superscriptaddress]{revtex4}
\usepackage{epsfig} 
\usepackage{amsmath} 
\usepackage{graphicx} 
\large

\def\gev{{\rm GeV}}

\def\bs{B \rightarrow  X_s\, \mu^+ \, \mu^-} 
\def\bd{B \rightarrow  X_d\, \mu^+ \, \mu^-} 
\def\bq{B \rightarrow  X_q\, \mu^+ \, \mu^-}

\def\dub{\delta_{ub}}
\def\dubp{\delta_{ub'}}
\def\dcbp{\delta_{cb'}}
\def\v44{V_{4 \times 4}}
\def\v34{V_{3 \times 4}}
\def\bea{\begin{eqnarray}}
\def\eea{\end{eqnarray}}

\def\barr{\begin{eqnarray}}
\def\earr{\end{eqnarray}}

\def\acp{A_{\rm CP}}
\def\V{\widetilde{V}}

\newcommand{\beq}{\begin{equation}}
\newcommand{\eeq}{\end{equation}}
\newcommand{\beqa}{\begin{eqnarray}}
\newcommand{\eeqa}{\end{eqnarray}}

\newcommand{\ltt}{\lambda_{t t'}^q}
\newcommand{\ltu}{\lambda_{tu}^q}

\newcommand{\re}{{\rm Re}}
\newcommand{\im}{{\rm Im}}

\newcommand{\spp}{\vphantom{\bigg(}}

\def\lesssim{\mathrel{\hbox{\rlap{\hbox{\lower4pt\hbox{$\sim$}}}\hbox{$<$}}}} 
\def\gtrsim{\mathrel{\hbox{\rlap{\hbox{\lower4pt\hbox{$\sim$}}}\hbox{$>$}}}} 

\begin{document}  
\title{\bf CP asymmetry in the decays 
$B \to (X_s, X_d) \, \mu^+ \, \mu^-$ \\with four generations}
 
\author{Ashutosh Kumar Alok}
\email{alok@theory.tifr.res.in}
\affiliation{Tata Institute of Fundamental Research, Homi Bhabha
Road, Mumbai 400005, India}

\author{Amol Dighe}
\email{amol@theory.tifr.res.in}
\affiliation{Tata Institute of Fundamental Research, Homi Bhabha
Road, Mumbai 400005, India}
     
\author{Shamayita Ray}
\email{shamayitar@theory.tifr.res.in}
\affiliation{Tata Institute of Fundamental Research, Homi Bhabha
Road, Mumbai 400005, India}

\date{\today} 

\preprint{TIFR/TH/08-42}

%\pacs{}
%\keywords{}

\begin{abstract}

We estimate the CP asymmetry $A_{\rm CP}(q^2)$ in the decays 
$B \to X_s\, \mu^+\, \mu^-$ and $B \to X_d\, \mu^+\, \mu^-$ in the
standard model (SM) with an additional fourth generation.
We use a parametrization that allows us to explore the complete
parameter space of the $4\times 4$ quark mixing matrix, and
constrain these parameters from the current data on $B$ decays.
We find that the enhancement in $A_{\rm CP}(q^2)$ depends strongly on 
the mass of the $t'$, the up-type quark in the fourth generation. 
For $m_{t'}$ around $400$ GeV, the CP asymmetry in the high-$q^2$ region 
($q^2 >  14.4\, \rm GeV^2$) can be enhanced by more than an order of magnitude 
for $B \to X_s\, \mu^+\, \mu^-$ and upto a factor of $6$ 
for $B \to X_d\, \mu^+\, \mu^-$. 
There is no enhancement in the low-$q^2$ region ($1<q^2<6 \,\rm GeV^2$). 
With increasing $m_{t'}$, $A_{\rm CP}(q^2)$ in the high-$q^2$ 
(low-$q^2$) region first decreases (increases) and then saturates
at a value a few times the SM prediction.
In the high-$q^2$ region of $\bs$, this saturation value may be up to 25 times the SM
expectation.

\end{abstract}

\maketitle 

\newpage
%%%%%%%%%%%%%%%%%%%%%%%%%%%%%%%%%%%%%%%%%%%%%%%%%%%%%%%%%%%%% 
\section{Introduction}
%%%%%%%%%%%%%%%%%%%%%%%%%%%%%%%%%%%%%%%%%%%%%%%%%%%%%%%%%%%%%%

Upcoming high statistics experiments at the LHC and 
Super-$B$ factories will provide stringent tests of the standard model 
(SM) via flavor physics involving $B$ decays. 
The large number of $B$ hadrons anticipated to be produced at 
these facilities will allow us to measure various flavor
changing neutral current (FCNC) interactions. 
The quark level FCNC transition $b \to s(d) l^+ l^-$, where $l=e,\mu,\tau$, are forbidden 
at the tree level in the SM and can occur only 
via one or more loops. Therefore they have the
potential to test higher order corrections to the SM and also to
constrain many of its possible extensions. The quark level FCNC transitions 
$b \to s(d) l^+ l^-$ give rise to the inclusive semileptonic decays 
$B \to X_s(X_d) \, l^+ \, l^-$.

It is always good to consider new physics effects in the 
observables which are either zero or highly suppressed in the SM. 
The reason is that any finite or large measurement of
such an observable will confirm the existence of new physics. 
The CP asymmetry in  $B \to (X_s, X_d) \, l^+ \, l^-$ is one such 
observable. 
The CP asymmetry in $B \to (X_s, X_d) \, l^+ \, l^-$ has been 
widely studied within the framework of the SM and its possible extensions
\cite{Du:1995ez,Ali:1998sf,Kruger:1996dt,Eygi:2003aj,Handoko:1997gd,Fukae:2001dm,Bashiry:2006wd}.
In the SM, the CP asymmetry in 
$B \to X_s\, l^+ \,l^-$ is $\sim 10^{-3}$ \cite{Du:1995ez,Ali:1998sf} 
whereas in $B \to X_d\, l^+ \,l^-$ it is 
$\sim\, (3-6)\%$ \cite{Ali:1998sf,Kruger:1996dt,Eygi:2003aj}.  
In the SM with three generations (SM3), 
the only source of CP violation 
is the unique phase in the Cabibbo-Kobayashi-Maskawa (CKM) 
quark mixing matrix.
However in many possible extensions of the SM, 
there can be extra phases contributing to the
CP asymmetry. 

In this paper we study the CP asymmetry in 
$B \to (X_s, X_d) \, \mu^+ \, \mu^-$ 
within the framework of the SM with an additional
fourth generation (SM4). 
There is no clear theoretical argument to restrict the number of
generations to three in the SM. 
Therefore in principle we can have four or more generations. 
The effects of the extra generation have been studied in the literature in detail 
\cite{London:1989vf,Hou:1989ty,Bamert:1994yq,Inami:1994nj,Masiero:1995fi,Novikov:1999af,Erler:1998ig,Hou:2005hd,Hou:2006jy,Kim:2007zzg,Soni:2008bc}.
The existence of new generation fermions that are
lighter than $M_Z/2 \approx 45$ GeV has been excluded by the data 
on the width of the $Z$ boson \cite{pdg}, whereas 
the existence of fermions heavier than
$M_Z \approx 91$ GeV has been excluded by the
existing data on the $Z$ boson parameters combined with the
masses of the $W$ boson and the top quark \cite{Maltoni:1999ta}.
However using the same data one can show that a few extra generations 
are possible provided the neutral leptons have masses around $50\, \rm GeV$
 \cite{Frampton:1999xi,SilvaMarcos:2002bz}.

The electroweak (EW) precision measurements impose severe constraints on the fourth
generation \cite{Maltoni:1999ta,Novikov:1994zg,Evans:1994an,He:2001tp,Novikov:2001md,Kribs:2007nz}.
 A considerable amount of fine tuning is required to accommodate a heavy
fourth generation top quark $t'$ ($m_{t'}>400\, \rm GeV$) in order not to violate the experimental constraints from 
the $S$ and $T$ parameters \cite{Kribs:2007nz}. The parameter space of fourth generation masses 
with minimal contributions to the EW precision oblique parameters, and 
in agreement with all
experimental constraints, is \cite{Kribs:2007nz}
\beqa
m_{l'}-m_{\nu'} &\simeq& (30 - 60)\, {\rm GeV} \nonumber \\
m_{t'}-m_{b'} &\simeq& \left(1+\frac{1}{5}\frac{m_H}{115\, {\rm GeV}}\right) \times {50\, \rm GeV}\;,
\eeqa
where $m_H$ is the Higgs mass and $m_{l'}, m_{\nu'}, m_{b'}$ are the 
masses of the fourth generation charged lepton $l'$,  neutrino $\nu'$ and the down type  
quark $b'$ respectively. We see that the EW precision data constrain the mass
splitting between $t'$ and $b'$ ($l'$ and $\nu'$) to be small, around $50\, {\rm GeV}$.

The fourth generation has a significant effect on the Higgs sector of the SM. For example,
 the $t'$ and $b'$ quarks  increase the effective $ggH$ coupling by a factor of roughly $3$ which will 
increase the production cross section $\sigma_{gg \to H}$ by almost an order of magnitude \cite{Gunion:1994zm,Gunion:1995tp}. 
The effect of the fourth generation
on Higgs physics has been studied in \cite{Kribs:2007nz,Arik:2001iw,Arik:2002ci,Arik:2005ed}. 
In \cite{Kribs:2007nz}, 
it was shown that in the SM4, Higgs masses between
$115-315\, (115-750)$ GeV are allowed by the EW precision data at the 68\% (95\%) C.L. 
Thus the EW precision data favor a heavy Higgs boson if the fourth generation is introduced.

Rare decays of $B$ mesons occur at loop level and hence they are 
sensitive to the generic extensions of the SM. The effects of the 
fourth generation on inclusive $B$ decays have been studied in the
literature \cite{Hou:1987hm,Huang:1999dk,Arhrib:2002md,Aliev:2003gp,solmaz:2003js}.
We employ the Dighe-Kim parametrization \cite{Kim:2007zzg}
of the $4\times 4$ quark mixing matrix (CKM4)
that allows us to treat the effects of the fourth generation perturbatively
and explore the complete parameter space available.
We generalize the notion of unitarity triangles to unitarity quadrilaterals, 
and calculate the CP asymmetry.

The paper is organized as follows. In Sec. \ref{effH}, we present 
the theoretical expressions for the decay rate and CP asymmetry in 
$B \to (X_s, X_d) \, \mu^+ \, \mu^-$. 
In Sec. \ref{ckm4sm4}, we study constraints on the elements
of CKM4, whereas in Sec. \ref{cpasys} and \ref{cpasyd} we present the estimates of CP asymmetry in $\bs$ and 
$\bd$ respectively.  Finally in Sec. \ref{concl}, we present our conclusions.

%%%%%%%%%%%%%%%%%%%%%%%%%%%%%%%%%%%%%%%%%%%%%%%%%%%%%%%%%%%%% 
\section{Decay rate and CP asymmetry in $B \to (X_s, X_d) \, \mu^+ \, \mu^-$}
\label{effH}
%%%%%%%%%%%%%%%%%%%%%%%%%%%%%%%%%%%%%%%%%%%%%%%%%%%%%%%%%%%%%
\subsection{Effective Hamiltonian and decay rate}

The effective Hamiltonian in the SM for the decay 
$b\rightarrow q \mu^+ \mu^-$, where $q=s,d$, may be written as
\begin{eqnarray}
H_{eff} &=& \frac{4G_{F}}{\sqrt{2}} V_{tb}^{*} V_{tq} \sum_{i=1}^{10}
C_{i}(\mu) O_{i}(\mu) \; ,
\end{eqnarray}
where the form of operators $O_i$  and the expressions for 
calculating the coefficients $C_i(\mu)$ are given in 
\cite{Buras:1994dj}.
The fourth generation only changes values of the Wilson coefficients 
$C_{7,8,9,10}$ via the virtual exchange of 
$t^\prime$.
The Wilson coefficients in the SM4 can be written as
\begin{equation}
C^{\rm tot}_{i}(\mu_b)=C_{i}(\mu_b)
+\frac{V^{*}_{t^{'}b}V_{t^{'}q}}{V^{*}_{tb}V_{tq}}C^{t'}_{i}(\mu_b),
\end{equation}
where $i=7,8,9,10$. The new Wilson coefficients $C^{t'}_{i}(\mu_b)$ 
can easily be calculated by substituting $m_{t^{'}} $ for $m_t$
in the SM3 expressions involving the $t$ quark.

The amplitude for the decay $B \rightarrow X_{q}\, \mu^{+}\, \mu^{-}$ 
in the SM4 is given by
\begin{eqnarray}
M &=& \frac{G_F \alpha}{\sqrt{2} \pi} V_{tb}^{*} V_{tq} 
      \bigg[C_9^{\rm tot}\, {\bar s_L}\gamma_{\mu}b_L\,{\bar \mu}\gamma^{\mu}\mu + 
    C^{\rm tot}_{10}\, {\bar s_L}\gamma_{\mu}b_L\,{\bar \mu}\gamma^{\mu}\gamma^5 \mu 
      \nonumber \\  
    & & \phantom{space here al} 
+ 2C^{\rm tot}_7\,m_b{\bar s_L}i\sigma_{\mu\nu}\frac{q^{\mu}}{q^2}b_R\,{\bar \mu}
    \gamma^{\nu}\mu \bigg],
\label{matrix}
\end{eqnarray}
where the Wilson coefficients are evaluated at $\mu_b$=$m_b$. 
The calculation of the differential decay rate gives
\begin{equation}
\frac{{\rm d}B(\bq)}{{\rm d}z}
 = \frac{\alpha^2 B(B\rightarrow X_c e {\bar \nu})}
 {4 \pi^2 f(\hat{m_c})\kappa(\hat{m}_c)} (1-z)^2\left(1-\frac{4t^2}{z}\right)^{1/2}
 \frac{|V_{tb}^{*}V_{tq}|^2}{|V_{cb}|^2} D(z)\,,
\end{equation}
where
 \begin{eqnarray}
 D(z) &=& |C_9^{\rm tot}|^2\left(1+\frac{2t^2}{z}\right)(1+2z)
      + 4|C_7^{\rm tot}|^2\left(1+ \frac{2t^2}{z}\right)\left(1+\frac{2}{z}\right) \nonumber \\
    & &  + |C_{10}^{\rm tot}|^2 \left[ ( 1 + 2z) + \frac{2t^2}{z}(1-4z)\right]
      +12 {\rm Re}(C_7^{\rm tot} C_{9}^{\rm tot*})\left(1+\frac{2t^2}{z}\right)\;.
\end{eqnarray}
Here $z \equiv q^2/m_b^2$, $t \equiv m_{\mu}/m_{b}$ and 
$\hat{m}_q=m_q/m_b$ for all quarks $q$. 
The phase space factor $f(\hat{m_c})$ in $B(B \to X_c e {\bar \nu})$ 
is given by \cite{Nir:1989rm}
\begin{equation}
f(\hat{m}_c) = 1 - 8\hat{m}^2_c + 8\hat{m}_c^6 - \hat{m}_c^8 - 
24\hat{m}_c^4 \ln \hat{m}_c \;.
\end{equation}
$\kappa(\hat{m_c})$ is the $1$-loop QCD correction factor 
\cite{Nir:1989rm}
 \begin{equation}
\kappa(\hat{m_c})=1-\frac{2\alpha_s(m_b)}{3\pi}\left[\left(\pi^2-\frac{31}{4}\right)(1-\hat{m_c})^2+\frac{3}{2}\right]\;.
\end{equation}
Within the SM3, the Wilson coefficients $C_7$ and $C_{10}$ are real. However the Wilson coefficient $C_9$ becomes slightly
complex due to the non-negligible terms induced by the continuum part of $u\bar{u}$ and $c\bar{c}$ loops 
proportional to 
$V_{ub}^*V_{uq}$ and $V_{cb}^*V_{cq}$, respectively. 
This complex nature of $C_9$ gives rise to
the CP asymmetry in $B \to (X_s, X_d) \, \mu^+ \, \mu^-$ in the SM3. 

In the framework of the SM4, the Wilson coefficients $C_{7}^{\rm tot }$, $C_{9}^{\rm tot}$, and $C_{10}^{\rm tot}$ are given by
\begin{eqnarray}
C_{7}^{\rm tot }& = & C_{7}(m_b)\,+\, \lambda_{tt'}^q\,C_{7}^{t' }(m_b)\;, \\
C_{9}^{\rm tot }& = & \xi_1\,+\, \lambda_{tu}^q\xi_2\,+\, 
\lambda_{tt'}^q\,C_{9}^{t' }(m_b)\;, \\
C_{10}^{\rm tot } & = & C_{10}(m_b)+ \lambda_{tt'}^q\,C_{10}^{t' }(m_b)\;,
\label{c10sm4}
\end{eqnarray}
where
\begin{equation}
\lambda_{tu}^q=\frac{\lambda_u^q}{\lambda_t^q}=\frac{V_{ub}^*V_{uq}}{V_{tb}^*V_{tq}}\;,
\end{equation}
\begin{equation}
\lambda_{tt'}^q=\frac{\lambda_{t'}^q}{\lambda_t^q}=\frac{V_{t'b}^*V_{t'q}}{V_{tb}^*V_{tq}}\;,
\end{equation}
so that all three relevant Wilson coefficients are complex in general.
The parameters $\xi_i$ are given by \cite{Buras:1994dj}
\begin{eqnarray}
\xi_1 & = & C_9(m_b) \, +\, 0.138 \,\omega(z)\,+\,g(\hat{m}_{c},z) (3 C_1 + C_2 + 3 C_3 +
C_4 + 3 C_5 + C_6)\nonumber\\&&- \frac{1}{2}g(\hat{m}_{d},z)
(C_3 + 3C_4) - \frac{1}{2}
   g(\hat{m}_{b},z)(4 C_3 + 4 C_4 + 3C_5 + C_6) \nonumber\\
  & & +\frac{2}{9} (3 C_3 + C_4 + 3C_5 + C_6)\;, \\
\xi_2 & = & [ g(\hat{m}_{c},z)- g(\hat{m}_{u},z)](3C_1 + C_2)\; .
\end{eqnarray}
Here
\bea 
\omega(z) & = & - \frac{2}{9} \pi^2 - \frac{4}{3}\mbox{Li}_2(z) - \frac{2}{3}
\ln z \ln(1-z) - \frac{5+4z}{3(1+2z)}\ln(1-z)  \nonumber \\
& &  - \frac{2 z (1+z) (1-2z)}{3(1-z)^2 (1+2z)} \ln z + \frac{5+9z-6z^2}{6
(1-z) (1+2z)} \; ,
\eea
with
\begin{equation}
\mbox{Li}_2(z)\,=\,-\int_0^t dt\, \frac{{\rm ln}(1-t)}{t}\;.
\end{equation}
The function $g(\hat m,z)$ represents the one loop
corrections to the four-quark operators $O_1-O_6$ and is given by \cite{Buras:1994dj}
\begin{eqnarray}
g(\hat m, z) &  = & -\frac{8}{9}\ln\frac{m_b}{\mu_b} - \frac{8}{9}\ln \hat m +
\frac{8}{27} + \frac{4}{9} x \\
& & - \frac{2}{9} (2+x) |1-x|^{1/2} 
\left\{\begin{array}{ll}
\left( \ln\left| \frac{\sqrt{1-x} + 1}{\sqrt{1-x} - 1}\right| - i\pi \right), &
\mbox{for } x \equiv \frac{4\hat m^2}{z} < 1 \nonumber \\
2 \arctan \frac{1}{\sqrt{x-1}}, & \mbox{for } x \equiv \frac
{4\hat m^2}{z} > 1,
\end{array}
\right. 
\end{eqnarray}
For light quarks, we have $\hat{m}_{u}\simeq \hat{m}_{d}\simeq0$. 
In this limit,
\beq
g(0, z) =  \frac{8}{27} -\frac{8}{9} \ln\frac{m_b}{\mu_b} - \frac{4}{9} \ln
z + \frac{4}{9} i\pi\;.
\eeq
We compute $g(\hat{m},z)$ at $\mu_b = m_b$.

%%%%%%%%%%%%%%%%%%%%%%%%%%%%%%%
\subsection{CP asymmetry in $\bq$}
\label{acpbq}
%%%%%%%%%%%%%%%%%%%%%%%%%%%%%%%

The CP asymmetry in $\bq$ is defined as
\begin{equation}
A_{\rm CP}(z)=\frac{(dB/dz)-(d\overline{B}/dz)}{(dB/dz)+(d\overline{B}/dz)}=
\frac{D(z)-\overline{D(z)}}{D(z)+\overline{D(z)}}\;,
\end{equation}
where $B$  and $\overline{B}$ represents the branching ratio of $\bar{B} \to  X_{q}l^+ l^-$
and its complex conjugate $B \to  \bar{X_{q}} l^+ l^- $ respectively. 
$d\overline{B}/dz$ can be obtained from $dB/dz$
by making the following replacements: 
\begin{eqnarray}
C_{7}^{\rm tot } =C_{7}(m_b)\,+\, \lambda_{tt'}^q\,C_{7}^{t' }(m_b)
& \to & \overline {C_{7}^{\rm tot }}=C_{7}(m_b)\,+\, 
\lambda_{tt'}^{q*}\,C_{7}^{t' }(m_b)\;, \\
C_{9}^{\rm tot }=\xi_1\,+\, \lambda_{tu}^q\xi_2\,+\, \lambda_{tt'}^q\,C_{9}^{t' }(m_b)
& \to &
\overline {C_{9}^{\rm tot }}=\xi_1\,+\, \lambda_{tu}^{q*}\xi_2\,+\, \lambda_{tt'}^{q*}\,C_{9}^{t' }(m_b)\;, \\
C_{10}^{\rm tot }=C_{10}(m_b)+ \lambda_{tt'}^q\,C_{10}^{t' }(m_b)
& \to & 
\overline {C_{10}^{\rm tot }}=C_{10}(m_b)+ \lambda_{tt'}^{q*}\,C_{10}^{t' }(m_b)\;.
\end{eqnarray}
Then
\beqa
D(z) - \overline{D(z)} 
&=& 2 \left( 1 + \frac{2 t^2}{z} \right) 
\bigg[ \im(\ltu) \left\{ 2(1+2z) \im(\xi_1 \xi_2^\ast) - 12 C_7 \im(\xi_2) \right\}  \nonumber \\
&& \phantom{2 \left( 1 + \frac{2 t^2}{z} \right) } 
+  X_{im} \left\{ (1+2 z) C_9^{t'} + 6 C_{7}^{t'} \right\} \bigg] \; ,
\label{acp_num} \\
D(z) + \overline{D(z)} 
&=& \left( 1 + \frac{2 t^2}{z} \right) 
\Bigl[ (1 + 2z) \left\{ B_1 + 2 C_9^{t'} \left( | \ltt |^2 C_9^{t'} + X_{re} \right) \right\} \Bigr. \nonumber \\
&& \Bigl.  + 12 \left\{ B_2 + 2 C_7 C_9^{t'} \re(\ltt) 
+ C_{7}^{t'} \left(  2 | \ltt |^2 C_9^{t'}  + X_{re} \right)  \right\}\Bigr] \nonumber \\
&&+  8  \left( 1 + \frac{2 t^2}{z} \right)  \left( 1 + \frac{2}{z} \right) |C_7^{\rm tot}|^2   \nonumber \\
&&+2 \left[ \left( 1 + 2 z \right) +  \frac{2 t^2}{z} \left( 1 - 4 z \right) \right] |C_{10}^{\rm tot}|^2 \; , 
\label{acp_den}
\eeqa
where
%2
\beqa
X_{re} &=& 2  \left\{ \re\left( \ltt \right) \re\left( \xi_1 \right) 
+ \re\left( \ltt {\ltu}^\ast \right) \re\left( \xi_2 \right)\right\} \; , \\
X_{im} &=& 2  \left\{ \im\left( \ltt \right) \im\left( \xi_1 \right) 
+ \im\left( \ltt {\ltu}^\ast \right) \im\left( \xi_2 \right)\right\} \; , 
\label{xim}\\
B_1 &=& 2 \left\{  |\xi_1|^2 + {|\ltu \xi_2|}^2 + 2 \re\left( \ltu \right) \re\left( \xi_1 \xi_2^\ast \right) \right\} \; , \\
B_2 &=& 2 C_{7} \left\{ \re(\xi_1) + \re(\ltu) \re(\xi_2)  \right\} \; , \\
|C_{10}^{\rm tot}|^2 
&=& {\left( C_{10} \right)}^2
+ |\ltt |^2 {\left( C_{10}^{t'} \right)}^2 + 2 C_{10} C_{10}^{t'} \re\left( \ltt \right) \; , \\
|C_{7}^{\rm tot}|^2 
&=&{\left( C_{7} \right)}^2 
+ |\ltt |^2 { \left( C_{7}^{t'} \right)}^2 
+ 2 C_{7}  C_{7}^{t'}  \re\left( \ltt \right) \; .
\eeqa

The theoretical calculations shown above for the branching ratio of
$\bq$ are rather uncertain in the intermediate $q^2$ region
($7$~GeV$^2 < q^2 < 12$~GeV$^2$) owing to the vicinity of
charmed resonances. The predictions are relatively more robust
in the lower and higher $q^2$ regions.
We therefore concentrate on calculating $\acp(q^2)$ 
in the low-$q^2$ ($1 \,{\rm GeV^2} < q^2 < 6\, {\rm GeV^2}$) 
and the high-$q^2$ ($14.4\, {\rm GeV^2} < q^2 < m_b^2$) regions. 
In terms of the dimensionless parameter $z=q^2/m_b^2$, 
the low-$q^2$ region corresponds to $0.043<z<0.26$ whereas 
the high $q^2$ region corresponds to $0.62<z<1$.

In order to estimate $\acp$, we need to know the magnitude and 
phase of $\lambda_{tu}^q$ and $\lambda_{tt'}^q$.
For this we use the Dighe-Kim (DK) parametrization of the 
CKM4 matrix elements, introduced in \cite{Kim:2007zzg}.

%%%%%%%%%%%%%%%%%%%%%%%%%%%%%%%%%%%%%%%%%%%%%%%
\section{The quark mixing matrix in SM4}
\label{ckm4sm4}
%%%%%%%%%%%%%%%%%%%%%%%%%%%%%%%%%%%%%%%%%%%%%%%%

\subsection{DK parametrization for the $4\times 4$ matrix CKM4}

The Cabibbo-Kobayashi-Maskawa (CKM) matrix in the SM is a $3\times3$ unitary matrix represented as
\begin{equation}
V_{\rm CKM3}=\left(\begin{array}{ccc}
V_{ud}& V_{us}&V_{ub}\\ V_{cd}&V_{cs}& V_{cb}\\
V_{td}&V_{ts}&V_{tb}
\end{array}\right)
\end{equation}

In the SM4, a general CKM matrix can be written as
follows:
\begin{equation}\label{standard}
V_{\rm CKM4}=\left(\begin{array}{cccc}
\V_{ud}& \V_{us}&\V_{ub}&\V_{ub'}\\ \V_{cd}&\V_{cs}& \V_{cb}&\V_{cb'}\\
\V_{td}&\V_{ts}&\V_{tb}&\V_{tb'}\\\V_{t'd}&\V_{t's}&\V_{t'b}&\V_{t'b'}
\end{array}\right)
\end{equation}
The above matrix can be described, with appropriate choices for the
quark phases, in terms of 6 real quantities and 3 phases.
The DK parametrization defines
\beq
\begin{tabular}{lll}
$\V_{us}  \equiv  \lambda$ , &
$\V_{cb}  \equiv A \lambda^2$ , &
$\V_{ub}  \equiv  A \lambda^3 C e^{-i\dub}$ , \\
$\V_{ub'} \equiv  p \lambda^3 e^{-i\dubp}$ , &
$\V_{cb'}  \equiv q \lambda^2 e^{-i\dcbp}$ , &
$\V_{tb'}  \equiv r \lambda $ .
\end{tabular}
\label{ckm4tab}
\eeq
The CKM4 matrix now looks like
\begin{equation}
%\label{standard}
V_{\rm CKM4}=\left(\begin{array}{cccc}
\phantom{sp}\# \phantom{sp} &\phantom{sp} \lambda \phantom{sp} &
A \lambda^3 C e^{-i\dub}&p \lambda^3 e^{-i\dubp}\\ 
 \# & \# & A \lambda^2&q \lambda^2 e^{-i\dcbp}\\
\# & \# & \# &r \lambda\\
\# & \# & \# & \# \\
\end{array}\right) \; .
\label{ckm4dk}
\end{equation}
The elements denoted by ``$\#$'' can be determined uniquely from the 
unitarity condition $V_{\rm CKM4}^\dagger V_{\rm CKM4} = I$ on CKM4. 
They can be calculated in the form of an expansion in the powers of 
$\lambda$ such that each element is accurate up to a multiplicative 
factor of $[1 + {\cal O}(\lambda^3)]$.
The matrix elements
$\V_{ud}$, $\V_{cd}$ and $\V_{cs}$ retain their SM3 values
\barr
\V_{ud} & = & 1 - \frac{\lambda^2}{2} + {\cal O}(\lambda^4) \label{vud} \; , \\
\V_{cd}  & = &  -\lambda + {\cal O}(\lambda^5) \label{vcd} \; , \\
\V_{cs}  & = &  1 - \frac{\lambda^2}{2}+ {\cal O}(\lambda^4)\;,
\earr
whereas the values of the matrix elements $V_{td}$, $V_{ts}$ and $V_{tb}$ 
are modified due to the presence of the additional quark generation:
\barr
\V_{td} & = & A \lambda^3 \left( 1 - C e^{i\dub} \right)
+ r \lambda^4 \left( q e^{i\dcbp} - p e^{i\dubp} \right) \nonumber \\
& & + \frac{A}{2} \lambda^5 \left( -r^2 + (C + C r^2) e^{i\dub} \right)
+ {\cal O}(\lambda^6) \; , \\
\V_{ts} & = & -A \lambda^2 - q r \lambda^3 e^{i\dcbp}
+ \frac{A}{2} \lambda^4 \left( 1 + r^2 - 2 C e^{i\dub} \right)
+ {\cal O}(\lambda^5) \; ,\\
\V_{tb} & = & 1 - \frac{r^2 \lambda^2}{2}+ {\cal O}(\lambda^4) \;.
\label{vtb}
\earr
In the limit $p=q=r=0$, only the elements present in the
$3\times 3$ CKM matrix retain nontrivial values, and the above expansion
corresponds to the Wolfenstein parametrization \cite{Wolfenstein} with
$C= \sqrt{\rho^2 + \eta^2}$ and $\dub = \tan^{-1}(\eta/\rho)$.
The remaining new CKM4 matrix elements are:
\barr
\V_{t'd} & = & \lambda^3 \left( q e^{i\dcbp} - p e^{i\dubp} \right)
+ A r  \lambda^4 \left( 1 + C e^{i\dub} \right) \nonumber \\
& & + \frac{\lambda^5}{2} \left(p e^{i\dubp}  - q r^2 e^{i\dcbp} +
pr^2   e^{i\dubp} \right)+ {\cal O}(\lambda^6) \; , \\
\V_{t's} & = & q \lambda^2  e^{i\dcbp} + A r \lambda^3 \nonumber \\
& & + \lambda^4 \left( -p e^{i\dubp} + \frac{q}{2} e^{i\dcbp} +
\frac{q r^2}{2} e^{i\dcbp} \right)+ {\cal O}(\lambda^5) \; ,\\
\V_{t'b} & = & - r \lambda+ {\cal O}(\lambda^4) \; ,\\
\V_{t'b'} & = & 1 - \frac{r^2 \lambda^2}{2} + {\cal O}(\lambda^4) \; .
\earr

We already have strong direct bounds on the magnitudes of the elements
of the CKM3 matrix. 
From the direct measurements of $|\V_{us}| = |V_{us}|, 
|\V_{cb}| = |V_{cb}|$ and
$|\V_{ub}/\V_{cb}| = |V_{ub}/V_{cb}|$ \cite{pdg}, 
which do not assume the unitarity 
of the CKM matrix, one can derive \cite{Kim:2007zzg}
\beq
0.216 < \lambda < 0.223 ~,~
0.76 < A  < 0.90 ~,~
0.23 < C < 0.59~~
\label{lambda-a-c}
\eeq
at 90\% C.L.. Also, the phase $\delta_{ub}$ can be constrained through
the measurement of 
$\gamma \equiv  {\rm Arg}( - V_{ub}^* V_{ud})/(V_{cb}^* V_{cd})$
since from (\ref{ckm4tab}), (\ref{vud}) and (\ref{vcd}),
\beq
{\rm Arg}\left(- \frac{V_{ub}^* V_{ud}}{V_{cb}^* V_{cd}} \right)
\approx {\rm Arg}\left(- \frac{\V_{ub}^* \V_{ud}}{\V_{cb}^* \V_{cd}} \right)
\approx \delta_{ub} \; .
 \eeq
The value of $\delta_{ub}$ is therefore restricted to lie between
$(26^\circ$--$125^\circ)$ at 90\% C.L..

Direct bounds on $p$ and $q$ can be obtained by
combining the direct measurements of the magnitudes of the elements
in the first two rows with the unitarity constraints.
We get the 90\% C.L. bounds on $|\V_{ub'}|$ and $|\V_{cb'}|$ as
\beq
|\V_{ub'}| < 0.094 ~~,~~
|\V_{cb'}| < 0.147 ~~,
\label{bd-ubp}
\eeq
which correspond to $ p < 9.0~,~q < 3.05$.
In addition, a strong constraint is obtained on
the combination $X_{bb}^L \equiv (V_{\rm CKM4}^\dagger V_{\rm CKM4})_{bb}$
through the measurements involving $Z \to b\bar{b}$, which give
$X_{bb}^L = 0.996 \pm 0.005$ \cite{delaguila}.
This translates to $|\V_{t'b}| < 0.11 $ at 90\% C.L., which
corresponds to $r < 0.5 $.

The observables $\Delta M_{B_s}$, $\Delta M_{B_d}$, $B \to X_s \gamma$, 
$B \to X_s \,\mu^+ \,\mu^-$, and $\sin 2\beta$ are complicated 
functions of the CKM parameters 
$\lambda, A, C, p, q, r, \delta_{ub'}\,,\delta_{ub},\,{\rm and}\, \delta_{cb'}$. 
Hence we take care of the constraints on these parameters 
numerically, without giving the analytic expressions explicitly here.

\section{CP asymmetry in $\bs$}
\label{cpasys}

%%%%%%%%%%%%%%%%%%%%%%%%%%%%%%%%%%%%%%%%%%%%%%%%%%%%%%%%%%%%%%%%%%%%%%
\subsection{Unitarity quadrilateral relevant for $B \to X_s \,\mu^+ \,\mu^-$}
\label{utbs}
%%%%%%%%%%%%%%%%%%%%%%%%%%%%%%%%%%%%%%%%%%%%%%%%%%%%%%%%%%%%%%%%%%%%%%%%

\begin{figure}[h] 
\centering
\includegraphics[width=0.8\textwidth]{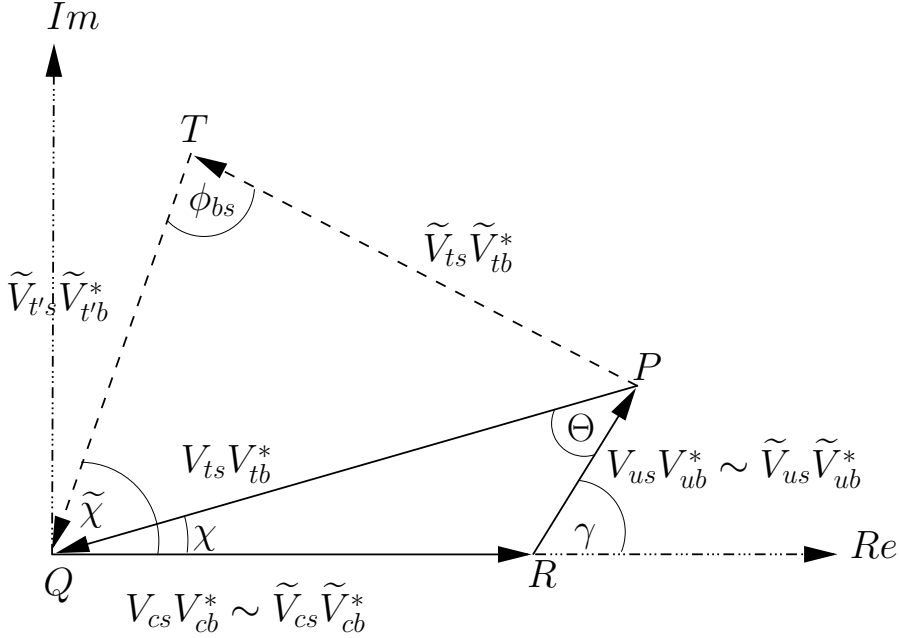} 
\caption{The "squashed" unitarity triangle (PQR) in the SM3  and the
corresponding unitarity quadrilateral (QRPT) in the SM4. 
\label{fig-tris}} 
\centering 
\end{figure}

 The ``squashed'' unitarity triangle in the SM3 that arises from the 
equation
\beq
 V_{cb}^*V_{cs}+V_{ub}^*V_{us}  + V_{tb}^*V_{ts} = 0 ~~.
\eeq
is shown in Fig.~\ref{fig-tris}.
The angles of this unitarity triangle are
\beq
\chi \equiv {\rm Arg}\left(-\frac{V_{cb}^* V_{cs}}{V_{tb}^* V_{ts}}
\right) ~,~
\Theta \equiv {\rm Arg}\left(-\frac{V_{tb}^* V_{ts}}{V_{ub}^* V_{us}}
\right) = \gamma - \chi ~,~
\pi - \Theta - \chi ~~.
\label{chi-def}
\eeq
The corresponding unitarity ``quadrilateral'' relation in the SM4 is
\beq
\V_{cb}^* \V_{cs} + \V_{ub}^* \V_{us} + \V_{tb}^* \V_{ts} +
\V_{t'b}^* \V_{t's}  = 0 ~~,
\eeq
This quadrilateral may be superimposed on the SM unitarity triangle
as shown in Fig.~\ref{fig-tris}.

%%%%%%%%%%%%%%%%%%%%%%
%%%%%%%%%%%%%%%%%%%%%%

The CP asymmetry in the SM3 depends on ${\rm Im}(\lambda^s_{tu})$,
as can be seen from eq.~(\ref{acp_num}). This quantity may
be written as
\beq
{\rm Im}(\lambda^s_{tu}) = - C \lambda^2 \sin \delta_{ub} 
+ {\cal O}(\lambda^3) \; ,
\eeq
which is the same as  the sine of the angle $\chi$ shown
in Fig.~\ref{fig-tris}.
With the introduction of the fourth generation, the contribution
to the CP asymmetry also comes from the quantity
${\rm Im}(\lambda^s_{tt'})$, which may be written as
\beq
{\rm Im}(\lambda^s_{tt'}) =
\frac{q r \sin \delta_{cb'}}{A} \lambda + {\cal O}(\lambda^2) \; ,
\eeq
which is the same as the sine of the angle $\widetilde\chi$ in
the figure.
Thus, the new CKM4 elements themselves tend to magnify the
CP violation by a factor of $\sim 1/\lambda \approx 5$.
There can of course be additional factors due to the
modified Wilson coefficients in SM4, which we will take care
of in our complete numerical analysis in the next section.

%%%%%%%%%%%%%%%%%%%%%%%%%%%%%%%%%%%%%%%%%%%%%%%%%%%%%%%%%%%%%%%%%
\subsection{Numerical calculation of $\acp(q^2)$ in $\bs$}
\label{numacps}
%%%%%%%%%%%%%%%%%%%%%%%%%%%%%%%%%%%%%%%%%%%%%%%%%%%%%%%%%%%%%%%%%%

In order to calculate $\acp(q^2)$ from the procedure outlined in 
Sec.~\ref{acpbq}, we need to know $\lambda_{tu}^q$ and $\lambda_{tt'}^q$. 
Using the DK parametrization, we have
\begin{eqnarray}
\lambda^s_{tt'}\,& = & \,\frac{e^{i\dcbp}qr\lambda}{A}\,+\,\left(r-\frac{e^{2i\dcbp}q^2r^2}{A^2}\right)\lambda^2 + {\cal O}(\lambda^3) \;, \\
\lambda_{tu}^s\,& = & \,- C e^{i\dub}\,\lambda^2 + {\cal O}(\lambda^3) \; .
\end{eqnarray}
Putting these values of $\lambda_{tu}^s$ and $\lambda_{tt'}^s$ in 
the relevant expressions in Sec.~\ref{acpbq},
we obtain $\acp(q^2)$ in $B\to X_s\, \mu^+\, \mu^-$.
The inputs used in the numerical analysis are shown in Table~\ref{tab1}. 

%%%%%%%%%%%%%%%%%%%%%%%%%%%%%%%%%%%%%%%%%%%
\begin{table} 
\begin{center} 
\begin{displaymath} 
\begin{tabular}{lcl} 
\hline 
\spp $G_F = 1.166 \times 10^{-5} \; \gev^{-2}$ & $\phantom{cc}$ & 
     $m_c/m_b=0.29\;$ \cite{ali-02}\\ 
\spp $\alpha = 1.0/129.0$ &  &
     $f_{B_s} \sqrt{ \hat{B_s}}=(0.270\pm0.030)\,\gev$ \cite{Blanke:2008ac}\\ 
\spp $\alpha_s(m_b)=0.220$ \cite{beneke-99} &  &
     $f_{B_d}\sqrt {\hat{B_d}}=(0.225\pm0.025)\,\gev$ \cite{Blanke:2008ac} \\ 
\spp $\tau_{B_s} = 1.45 \times 10^{-12}\; s$  &  &
     $\Delta m_{s}= (1.17 \pm 0.008) \times 10^{-11}\,\gev$\\ 
\spp $\tau_{B_d} = 1.53 \times 10^{-12}\; s$  &  &
     $\Delta m_{d}= (3.337 \pm 0.033) \times 10^{-13}\,\gev$ \\ 
\spp $m_{\mu}=0.105 \;\gev$ &  &
     $\sin2\beta=0.681\pm0.025$ \\ 
\spp $m_W= 80.40 \; \gev$ &  &
     $\delta_{ub}(\equiv \gamma)=\left(77^{+30}_{-32}\right)^{\circ}$\\ 
\spp $m_t= 172.5 \; \gev$ &  &
     $B(B \to X_c \ell \nu)=0.1061\pm0.0016\pm0.0006$ \cite{Aubert:2004aw} \\ 
\spp $m_b=4.80\; \gev $ \cite{ali-02} &  &
     $B(B \to X_s\, \mu^+\, \mu^-)_{q^2>14.4\,\gev}=(0.44\pm0.12)\times 10^{-6}$ \cite{Aubert:2004it,Iwasaki:2005sy}\\
\spp $m_{B_s}=5.366 \; \gev$ &  &
     $B(B \to X_s \, \gamma)=(3.55\pm0.25)\times 10^{-4}$ \cite{Barberio:2008fa}\\
\spp $ m_B=5.279 \; \gev$  &  &
     \\
     \hline 
\end{tabular} 
\end{displaymath} 
\caption{Numerical inputs used in our 
  analysis. Unless explicitly specified, they are taken from the 
  Review of Particle Physics~\cite{pdg}.\label{tab1}} 
\end{center} 
\end{table} 
%%%%%%%%%%%%%%%%%%%%%%%%%%%%%%%%%%%%%%%%%%%%%%%%%%%%%%%%

%%%%%%%%%%%%%%%%%%%%%%%%%%%%%%%%%%%%%%
\begin{figure}
\hbox{\hspace{0.002cm}
\hbox{\includegraphics[scale=0.7]{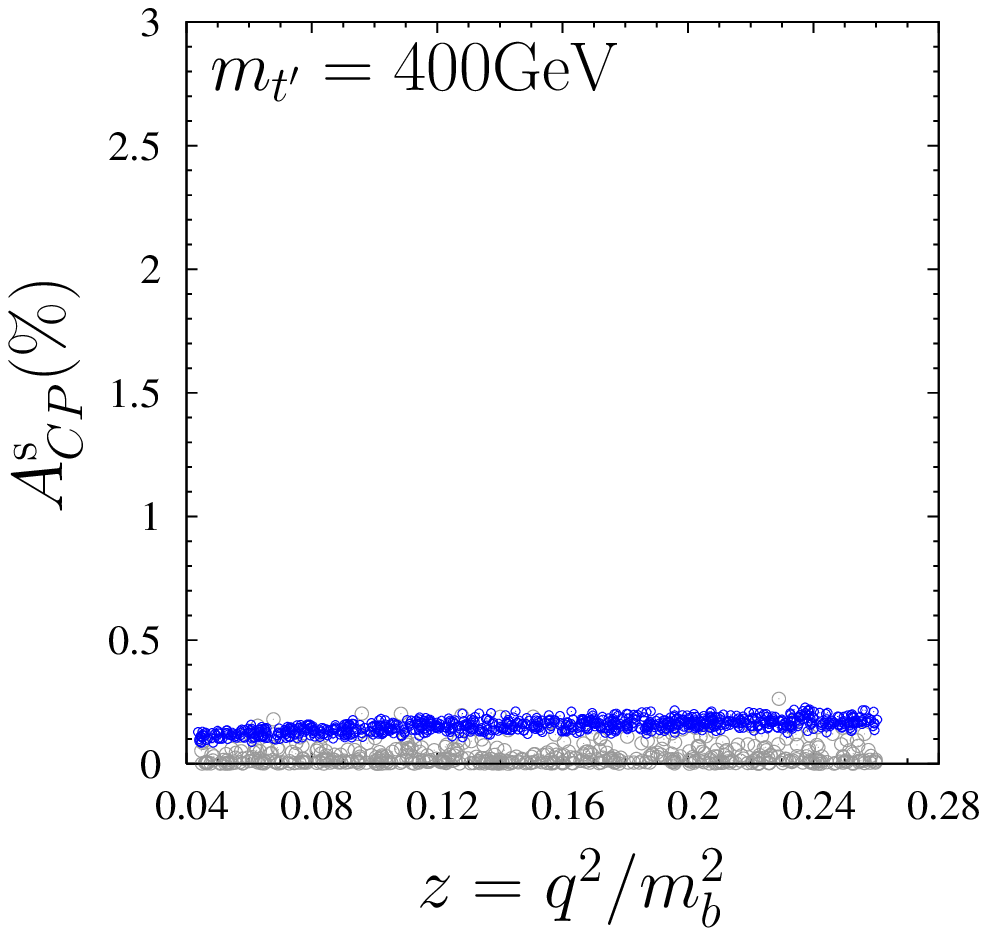}}
\hspace{-0.08cm}
\hbox{\includegraphics[scale=0.7]{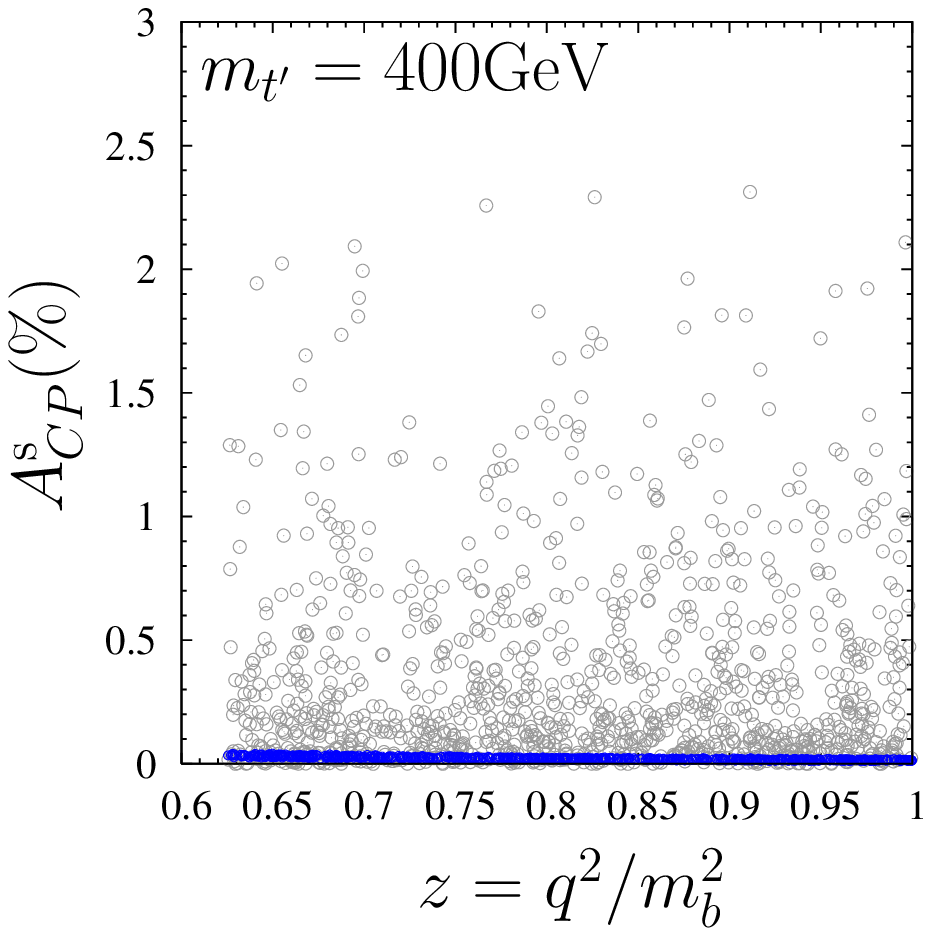}}}
\hbox{\hspace{0.002cm}
\hbox{\includegraphics[scale=0.7]{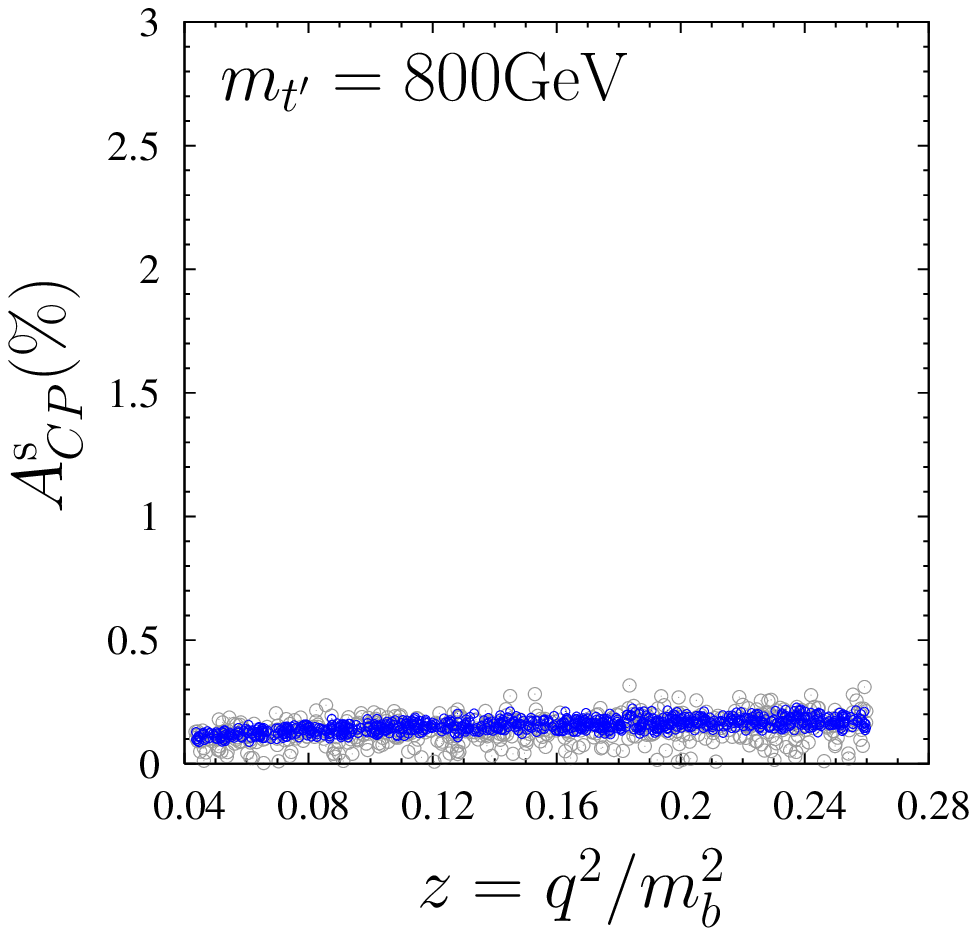}}
\hspace{-0.08cm}
\hbox{\includegraphics[scale=0.7]{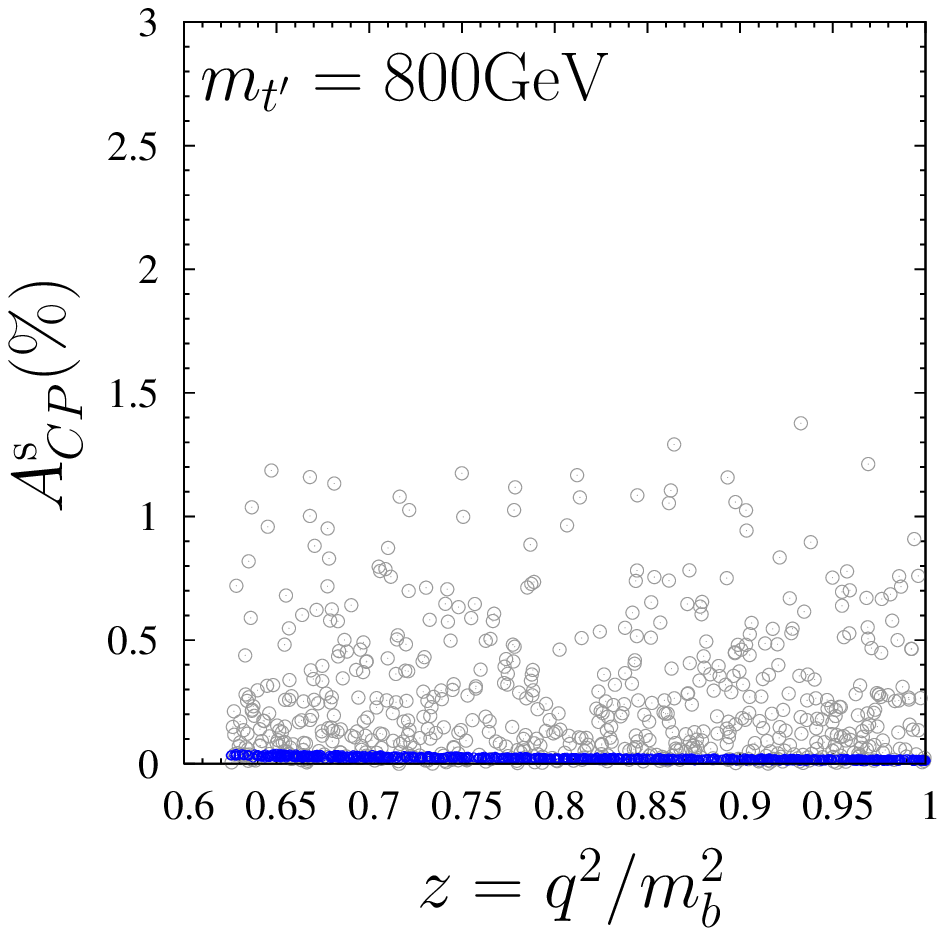}}}
\hbox{\hspace{0.002cm}
\hbox{\includegraphics[scale=0.7]{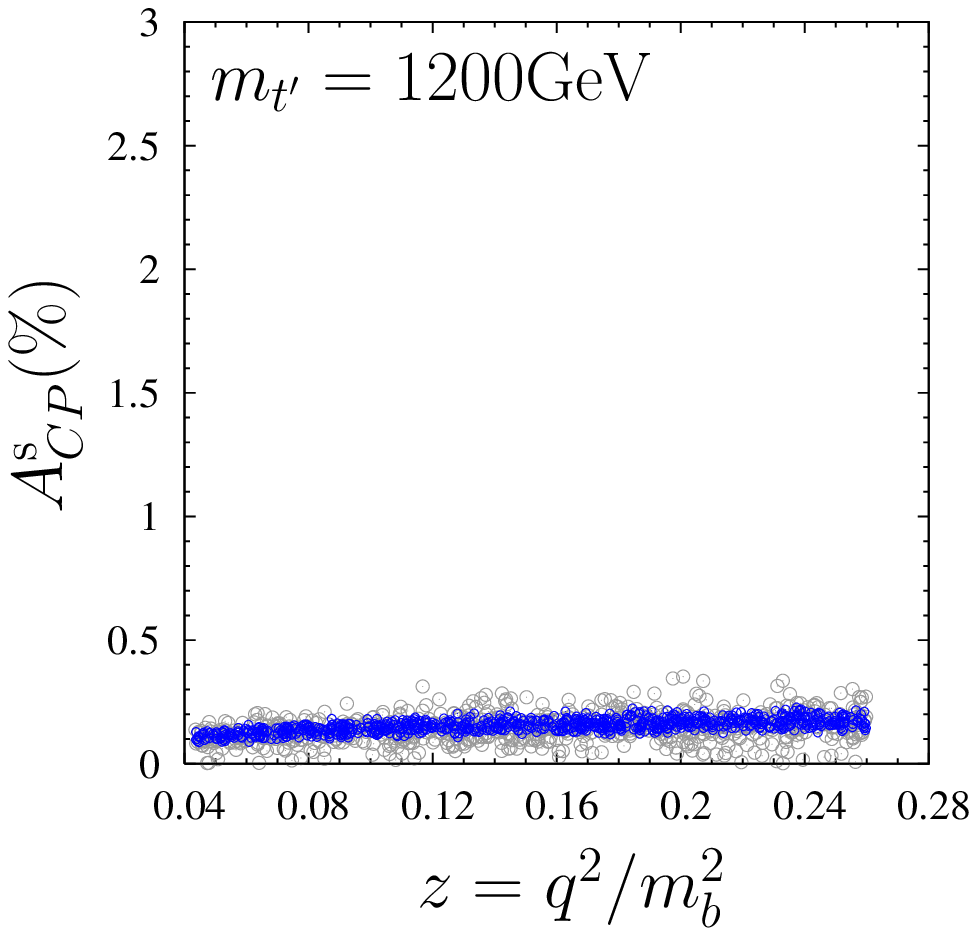}}
\hspace{-0.08cm}
\hbox{\includegraphics[scale=0.7]{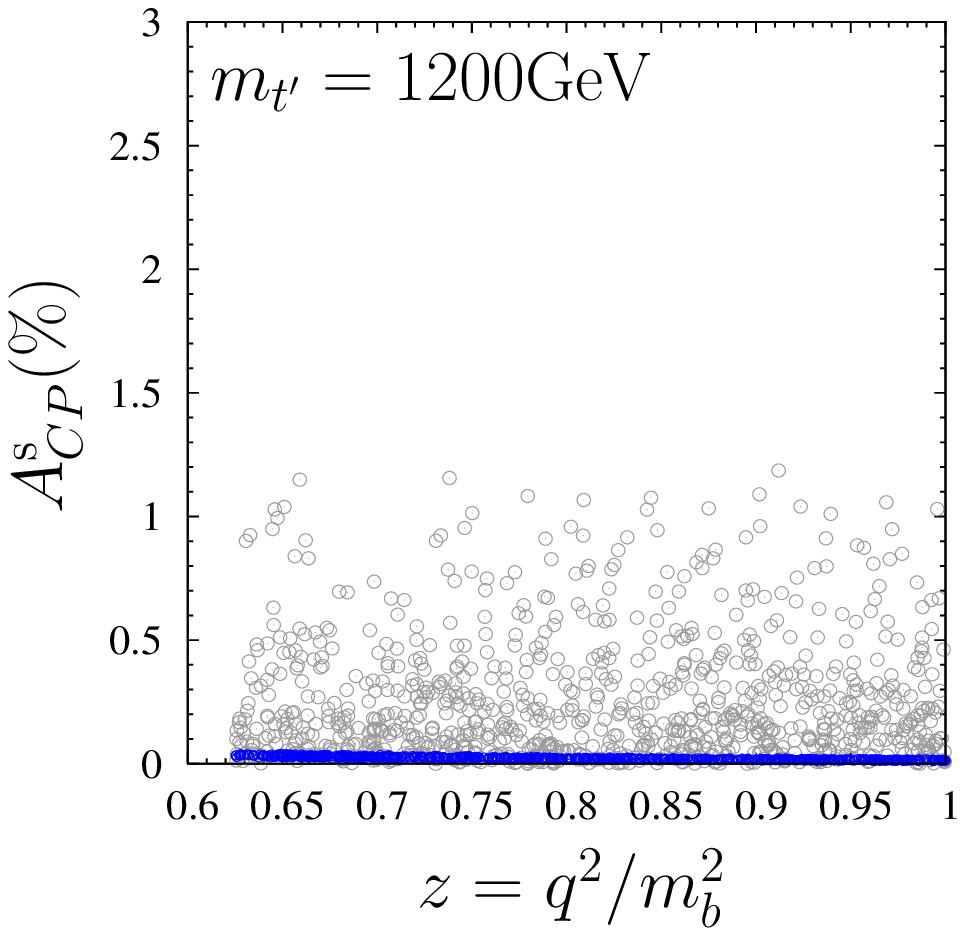}}}
\caption{ $\acp(z)$ vs $z$ plot in the low-$q^2$ (left panel) and the 
high-$q^2$ (right panel) regions for the decay
$B\to X_s\, \mu^+\, \mu^-$ for $m_{t'}=(400,\,800,\,1200)\,\rm GeV$.
The blue band represents the SM3 prediction whereas the 
grey circles correspond to the possible values that can be obtained
in the SM4.
\label{acps01}}
\end{figure}

%%%%%%%%%%%%%%%%%%%%%%%%%%%%%%%%%%%%%

Fig.~\ref{acps01} shows $\acp(q^2)$ in the low and high $q^2$ regions 
for the decay $\bs$ for $m_{t'}=(400,\, 800,\, 1200)\,\rm GeV$. 
Clearly for $m_{t'}=400\,\rm GeV$, for most of the allowed regions 
of the parameter space, the SM4 prediction for $\acp(q^2)$
in the low-$q^2$ region is either below the SM3 prediction or 
consistent with it. 
However in the high-$q^2$ region, the SM4 prediction can be as high 
as $2.5\%$, which is about $40$ times the SM3 prediction. 
There is thus a significant enhancement in $\acp(q^2)$ in the
high-$q^2$ region. 

\begin{table}
\begin{center}
\begin{tabular}{cccccccccccccc}
\hline
&$\phantom{space}$ & \multicolumn{5}{c}{$[A_{\rm CP}^s(q^2)]_{\rm max}$ (low $q^2$)} 
& & \multicolumn{5}{c}{$[A_{\rm CP}^s(q^2)]_{\rm max}$  (high $q^2$ )} \\
$m_{t'}$ (GeV) 
& $\phantom{spc}$ & SM3 &$\phantom{spc}$ & SM4 &$\phantom{spc}$ & SM4/SM3  
& $\phantom{space}$ & SM3 &$\phantom{spc}$ & SM4 &$\phantom{spc}$ & SM4/SM3
\\
\hline 
$400$ & &$0.25\%$ & & $0.25\%$ & & $1.0$ & & $0.05\%$ & & $2.3\%$ & & $46$ \\
$800$ & & $0.25\%$  & & $0.3\%$  & & $1.2$ & & $0.05\%$ & & $1.4\%$ & & $28$\\
$1200$ & &$0.25\%$ & & $0.3\%$ & & $1.2$ & & $0.05\%$ & & $1.3\%$ & & $26$\\
\hline
\end{tabular}
\end{center}
\caption{Comparison of $\acp(q^2)$ in the SM3 and in the SM4 for $\bs$ 
at different $m_{t'}$ values}
\label{tab:s}
\end{table}

Table~\ref{tab:s} shows the ratio of the maximum $\acp(q^2)$ allowed
within the SM4 and that allowed in the SM3. It can be seen that with increasing
$m_{t'}$, the enhancement in $\acp(q^2)$ at low-$q^2$ (high-$q^2$) increases
(decreases) and then saturates at $\sim 1.2$ ($25$) times the SM value.
Thus, while the low-$q^2$ region is rather insensitive to the
effects of the fourth generation, the high-$q^2$ region may show a 
significant asymmetry that can easily be shown to be beyond the
limits of the SM3.

The saturation in $\acp(q^2)$ at large $m_{t'}$ may be understood as follows.
The Wilson coefficient $C_{10}$ becomes very large as compared to 
$C_7$ and $C_9$ for large $m_{t'}$. Hence from eq.~(\ref{c10sm4}), 
it is obvious that $\lambda_{tt'}^s$ must be very small for large 
$m_{t'}$ so as to keep the branching ratio within the experimental range. 
Hence in the limit of large $m_{t'}$, we have $\lambda_{tt'}^s \to 0$. 
In this limit, the $X_{im}$ term 
in eq.~(\ref{acp_num}) vanishes and the numerator of 
$\acp(q^2)$ becomes
\beqa
D(z) - \overline{D(z)} 
&=& 2 \left( 1 + \frac{2 t^2}{z} \right) 
\left[ \im(\ltu)\left\{ 2(1+2z) \im(\xi_1 \xi_2^\ast) 
- 12 C_7 \im(\xi_2)\right\} \right] \; .
\label{acp_num1}
\eeqa
The right hand side of eq.~(\ref{acp_num1}) has only a weak dependence 
on $m_{t'}$ and hence remains almost constant for large $m_{t'}$. 
$D(z) + \overline{D(z)}$, on the other hand, is just obtained from
the branching ratio of $\bs$, an experimentally measured value.
The ratio of these two quantities, $\acp(q^2)$, is therefore 
rather independent of $m_{t'}$ at large $m_{t'}$.
This fact is reflected in the $\acp$ plots: there is not much difference 
in the $\acp(q^2)$ prediction for $m_{t'}=800\,\rm GeV$ 
and $m_{t'}=1200\,\rm GeV$.

%%%%%%%%%%%%%%%%%%%%%%%%%%%%%%%%%%%%%%%%%%%%%%%%%%%%%%%%%%%%% 
\section{CP asymmetry in $\bd$}
\label{cpasyd}
%%%%%%%%%%%%%%%%%%%%%%%%%%%%%%%%%%%%%%%%%%%%%%%%%%%%%%%%%%%%%%

%%%%%%%%%%%%%%%%%%%%%%%%%%%%%%%%%%%%%%%%%%%%%%%%%%
\subsection{Unitarity quadrilateral relevant for $B \to X_d \,\mu^+ \,\mu^-$}
\label{utbd}
%%%%%%%%%%%%%%%%%%%%%%%%%%%%%%%%%%%%%%%%%%%%%%%%%%%%

\begin{figure}[h] 
\centering
\includegraphics[width=0.8\textwidth]{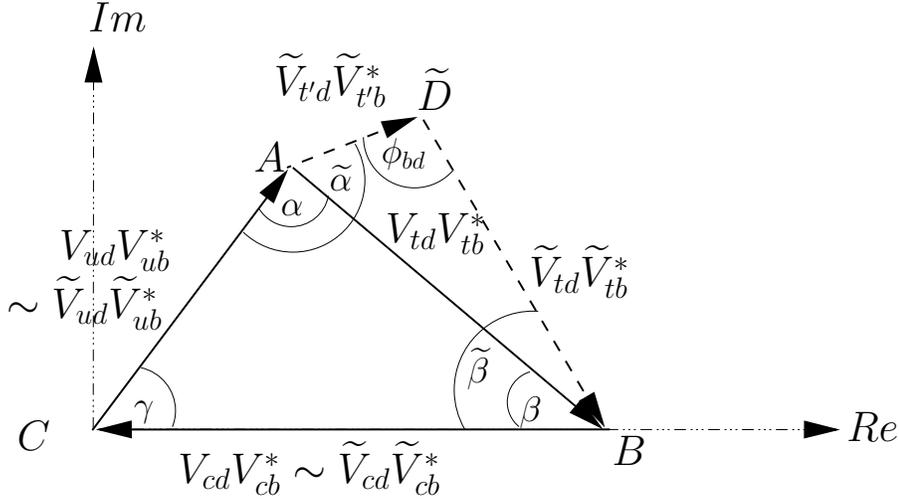} 
\caption{The unitarity triangle (ABC) in the SM3  and the
corresponding unitarity quadrilateral (ACBD) in the SM4. 
\label{fig-trid}} 
\centering 
\end{figure}

The ``standard'' unitarity triangle in the SM3,
which arises from the equation
\beq
V_{ub}^*V_{ud} + V_{cb}^*V_{cd} + V_{tb}^*V_{td} = 0 ~~,
\eeq
is shown in Fig.~\ref{fig-trid}
The angles of this unitarity triangle are defined as
\beq
\alpha \equiv {\rm Arg}\left(-\frac{V_{tb}^* V_{td}}{V_{ub}^* V_{ud}}
\right) ~,~
\beta \equiv {\rm Arg}\left(-\frac{V_{cb}^* V_{cd}}{V_{tb}^* V_{td}}
\right) ~,~
\gamma \equiv {\rm Arg}\left(-\frac{V_{ub}^* V_{ud}}{V_{cb}^* V_{cd}}
\right) ~.
\label{abg-def}
\eeq
The corresponding unitarity relation in the SM4 is
\beq
\V_{ub}^* \V_{ud} + \V_{cb}^* \V_{cd} + \V_{tb}^* \V_{td} +
\V_{t'b}^* \V_{t'd}  = 0 ~~,
\label{quad1}
\eeq
This quadrilateral
may be superimposed on the SM unitarity triangle as shown in
Fig.~\ref{fig-trid}.

%%%%%%%%%%%%%%%%%%%%%%
%%%%%%%%%%%%%%%%%%%%%%

The CP asymmetry in SM3 depends on ${\rm Im}(\lambda^d_{tu})$,
as can be seen from eq.~(\ref{acp_num}). This quantity may
be written as
\beq
{\rm Im}(\lambda^d_{tu}) = - {\rm Arg}\left(\frac{e^{i\delta_{ub}}}
{1 - C e^{i\delta_{ub}}}\right) + {\cal O}(\lambda) \; ,
\eeq
which is the same as  the sine of the angle $\beta$ shown
in Fig.~\ref{fig-trid}.
With the introduction of the fourth generation, contribution
to the CP asymmetry also comes from the quantity
${\rm Im}(\lambda^d_{tt'})$, which may be written as
\beq
{\rm Im}(\lambda^d_{tt'}) = {\cal O}(\lambda) \; .
\eeq
Thus, the additional contribution to the CP violation
from the complex nature of the CKM4 elements is rather small.
The enhancement in $\acp(q^2)$, if any, therefore has to come from
the modified values of the Wilson coefficients.
We calculate the enhancement numerically in the next section.

%%%%%%%%%%%%%%%%%%%%%%%%%%%%%%%%%%%%%%%%%%%%%%%%%%%%%%%%%%%%%
\subsection{Numerical calculation of $\acp(q^2)$ in $\bd$}
\label{numacpd}
%%%%%%%%%%%%%%%%%%%%%%%%%%%%%%%%%%%%%%%%%%%%%%%%%%%%%%%%%%%

We now consider $\lambda_{tu}^d$ and $\lambda_{tt'}^d$ for
the calculation of  $\acp(q^2)$ in $\bd$ using the procedure
outlined in Sec.~\ref{acpbq}.
Using the DK parametrization, we obtain
\bea
\lambda^d_{tt'} &=& \frac{\left(p e^{i\dubp}\,- qe^{i\dcbp}\right)r\lambda}{A\left(1-Ce^{i\dub}\right)}  + {\cal O}(\lambda^2)  \; ,\\
\label{ldttp}
\lambda_{tu}^d &=& \frac{e^{i\dub}}{1-Ce^{i\dub}}+\frac{e^{i\dub}\left(p e^{i\dubp}- q e^{i\dcbp} \right)r\lambda}{A\left(1-Ce^{i\dub}\right)^2} +
{\cal O}(\lambda^2) \; .
\eea
For our numerical analysis, we use the expressions correct
up to ${\cal O}(\lambda^2)$.

\begin{figure}
\hbox{\hspace{0.002cm}
\hbox{\includegraphics[scale=0.7]{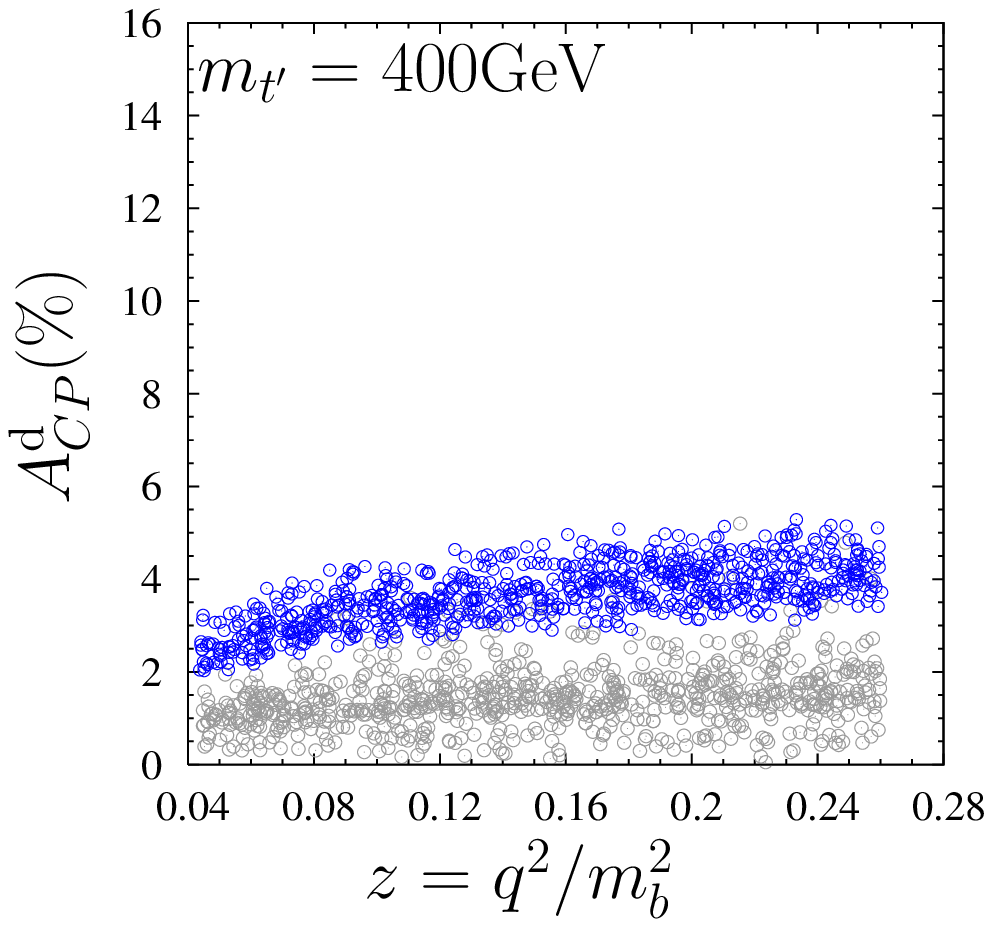}}
\hspace{-0.08cm}
\hbox{\includegraphics[scale=0.7]{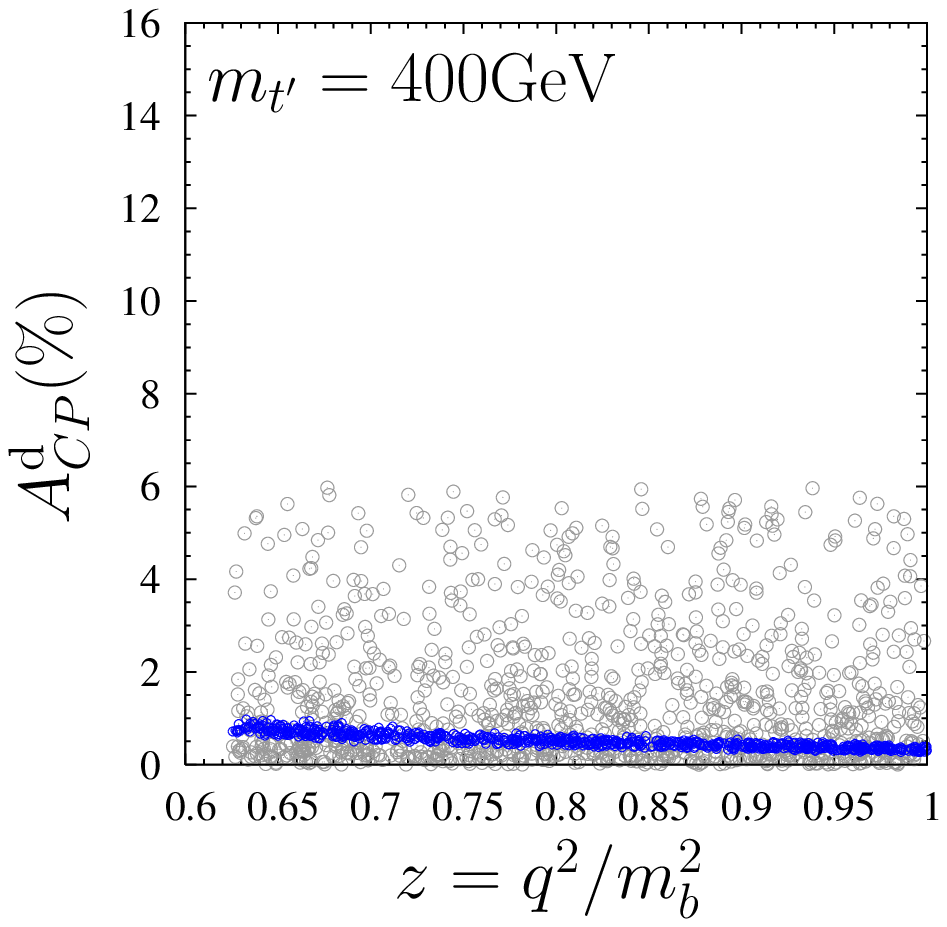}}}
\hbox{\hspace{0.002cm}
\hbox{\includegraphics[scale=0.7]{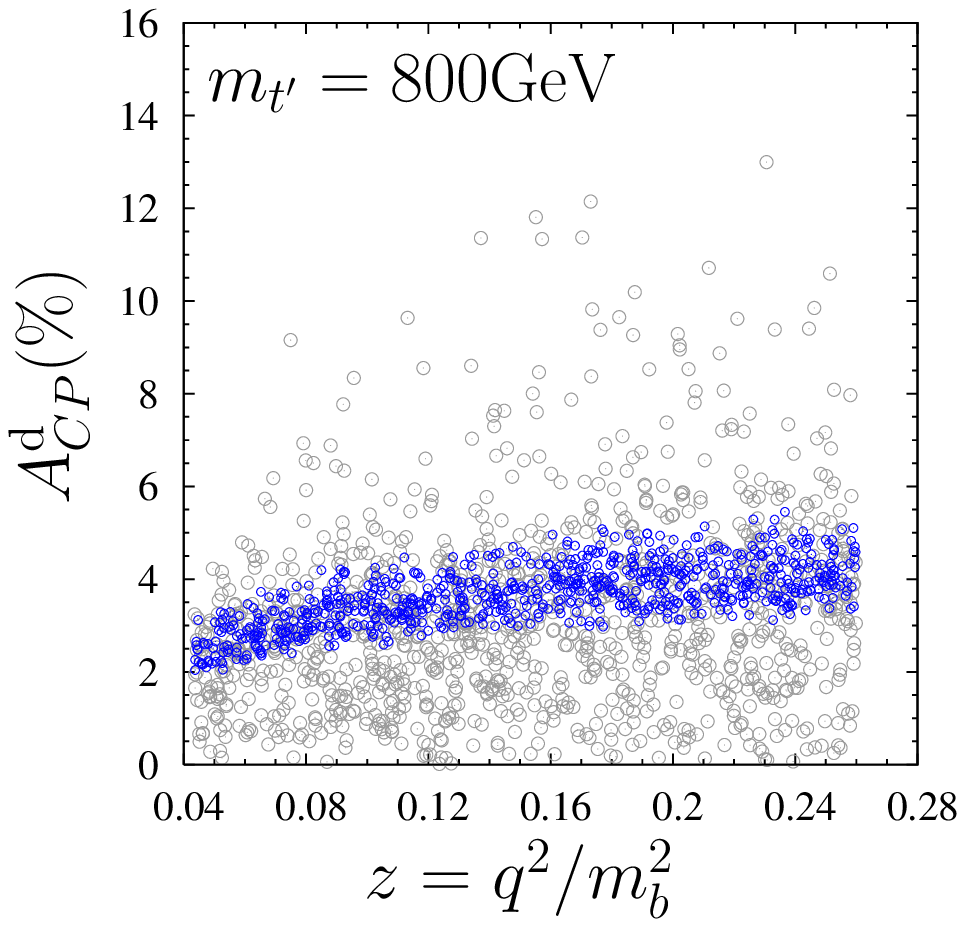}}
\hspace{-0.08cm}
\hbox{\includegraphics[scale=0.7]{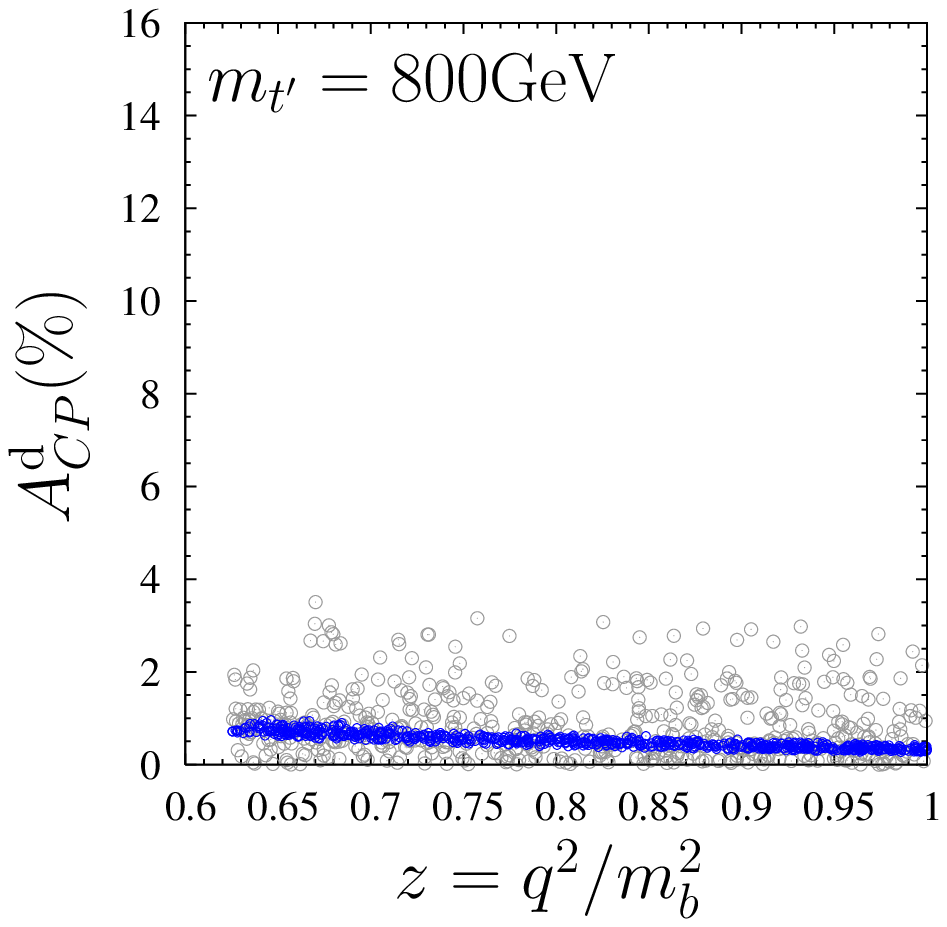}}}
\hbox{\hspace{0.002cm}
\hbox{\includegraphics[scale=0.7]{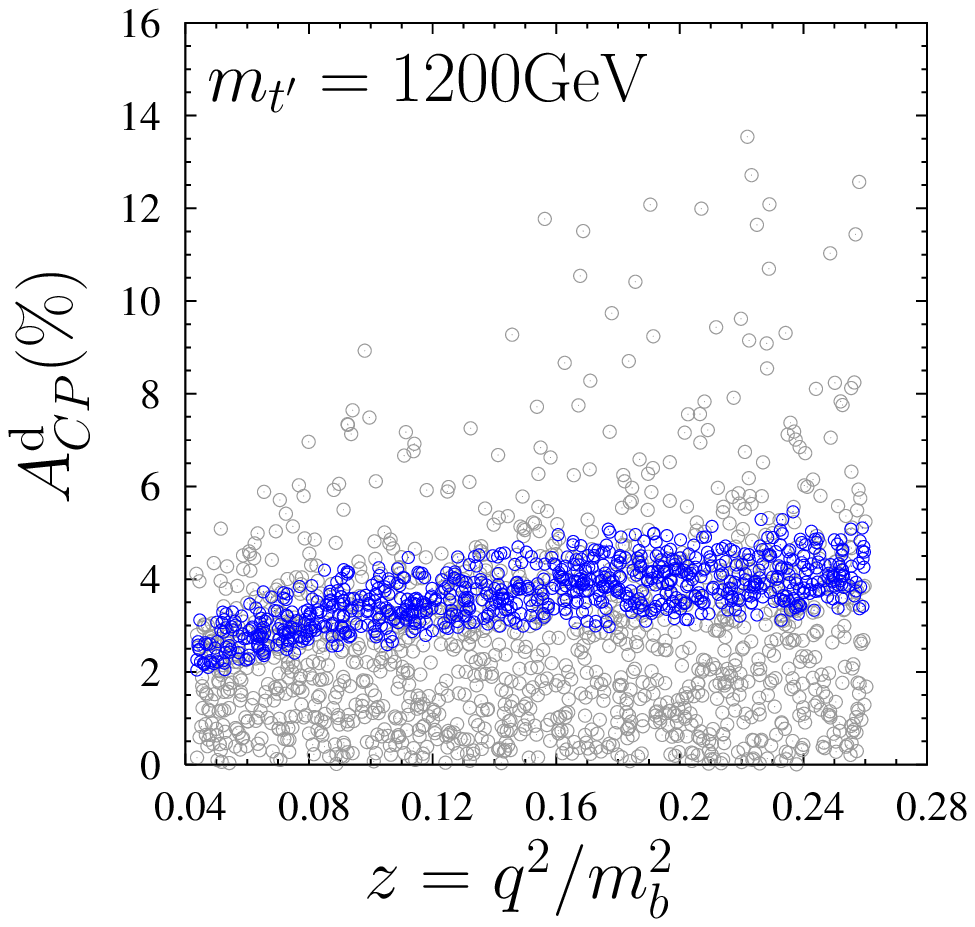}}
\hspace{-0.08cm}
\hbox{\includegraphics[scale=0.7]{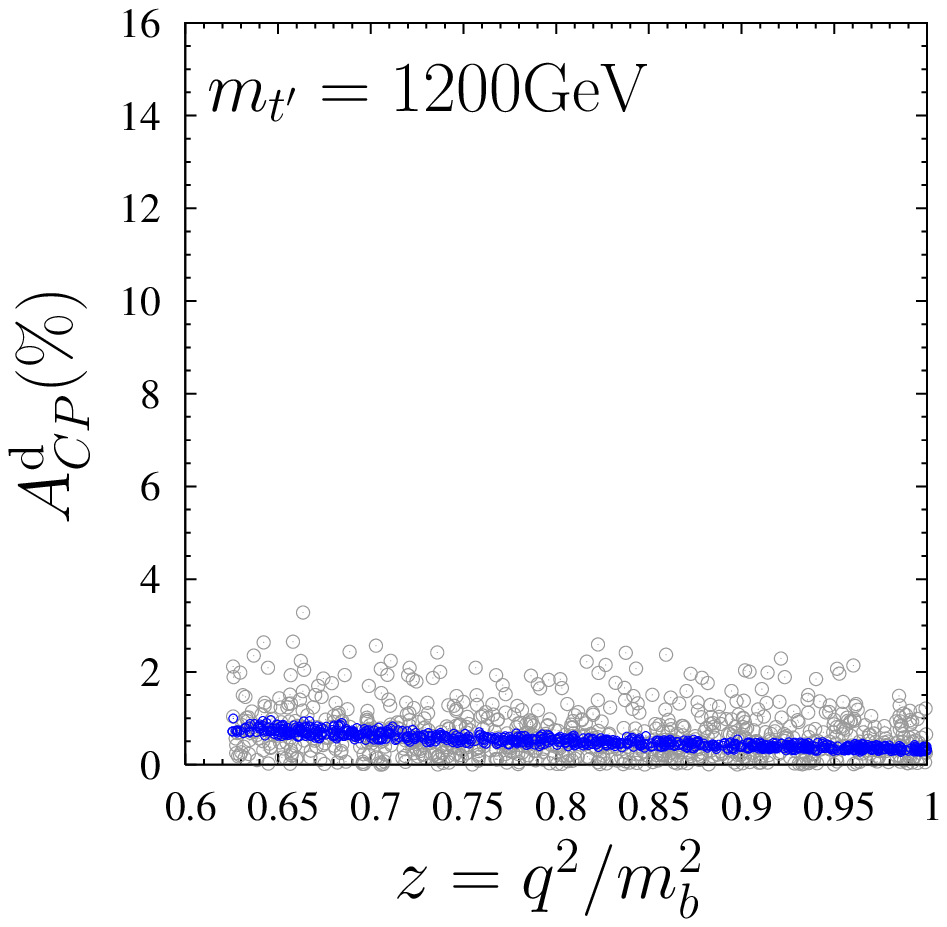}}}
\caption{ $\acp(z)$ vs $z$ plot in (a) the low-$q^2$ and (b) the high-$q^2$ 
region for the decay $B\to X_d\, \mu^+\, \mu^-$ for 
$m_{t'}=(400,\, 800,\, 1200)\,\rm GeV$.
\label{acpd01}}
\end{figure}

Fig.~\ref{acpd01} shows the  $\acp(q^2)$ distribution in the low-$q^2$ 
and the high-$q^2$ regions for $m_{t'}=(400,\, 800,\, 1200)\,\rm GeV$. 
Here we find that for $m_{t'}=400\,\rm GeV$, the low-$q^2$ prediction 
in the SM4 is either consistent with or below the SM3 prediction whereas 
in the high-$q^2$ region, the SM4 prediction can be as high as $6\%$,
which is about $6$ times the SM3 prediction. 
There is thus a significant enhancement in $\acp(q^2)$ in the high-$q^2$
region.

%%%%%%%%%%%%%%%%%%%%%%%%%%%%%%%%%%%%%%

%%%%%%%%%%%%%%%%%%%%%%%%%%%%%%%%%%%%%

Table~\ref{tab:d} shows the ratio of the maximal values of $\acp(q^2)$ allowed
within the SM4 and that allowed in the SM3. It can be seen that with increasing
$m_{t'}$, the enhancement in $\acp(q^2)$ at low-$q^2$ (high $q^2$) increases
(decreases) and then saturates at $\sim 2.5$ ($3$) times the SM3 value.
At low $m_{t'}$, the enhancement in the high-$q^2$ region is rather
large, and makes this region more suitable for the detection of
a deviation from the SM3 expectation, just like in the case of $\bs$.
However at high $m_{t'}$, the enhancement over the SM3 value is similar in 
both the regions, so that the higher branching ratio at low-$q^2$ 
and the higher value of $\acp(q^2)$  therein makes the analysis
of $\bd$ at low $q^2$ an interesting prospect.

\begin{table}
\begin{center}
\begin{tabular}{cccccccccccccc}
\hline
&$\phantom{space}$ & \multicolumn{5}{c}{$[A_{\rm CP}^d(q^2)]_{\rm max}$ (low $q^2$)} 
& & \multicolumn{5}{c}{$[A_{\rm CP}^d(q^2)]_{\rm max}$  (high $q^2$ )} \\
$m_{t'}$ (GeV) 
& $\phantom{spc}$ & SM3 &$\phantom{spc}$ & SM4 &$\phantom{spc}$ & SM4/SM3  
& $\phantom{space}$ & SM3 &$\phantom{spc}$ & SM4 &$\phantom{spc}$ & SM4/SM3
\\
\hline 
$400$ & &$5.5\%$ & & $5.5\%$ & & $1.0$ & & $1.0\%$ & & $6.0\%$ & & $6.0$ \\
$800$ & & $5.5\%$  & & $13.5\%$  & & $2.45$ & & $1.0\%$ & & $4.0\%$ & & $4.0$\\
$1200$ & &$5.5\%$ & & $13.5\%$ & & $2.45$ & & $1.0\%$ & & $3.0\%$ & & $3.0$\\
\hline
\end{tabular}
\end{center}
\caption{Comparison of $\acp(q^2)$ in the SM3 and in the SM4 for $\bd$ at 
different $m_{t'}$ values}
\label{tab:d}
\end{table}

The same arguments as given in Sec.~\ref{numacps} in the case of $\bs$
for the saturation of $\acp(q^2)$ at large $m_{t'}$ also apply to
$\bd$. The allowed range $\acp(q^2)$ at 800 GeV and 1200 GeV is 
then almost identical, as can be seen in Fig.~\ref{acpd01}.

%%%%%%%%%%%%%%%%%%%%%%%%%%%%%%%%%%%%%%%%%%%%%%%%%%%%%%%%%%%%% 
\section{Conclusions}
\label{concl}
%%%%%%%%%%%%%%%%%%%%%%%%%%%%%%%%%%%%%%%%%%%%%%%%%%%%%%%%%%%%%%

In this paper we study the CP asymmetry in the decays $\bs$ and $\bd$ in the
standard model with an additional fourth generation using the Dighe-Kim 
parametrization, which allows us to treat the problem as
a perturbative expansion in the Cabibbo angle $\lambda$,
and explore the complete parameter space of the
$4\times 4$ quark mixing matrix.
We use constraints from the present measurements of 
$\Delta M_{B_s}$, $\Delta M_{B_d}$, $\sin 2\beta$, and the branching
ratios of $B \to X_c e \bar{\nu}$, $B \to X_s \,\gamma$, $\bs$.
The results may be summarized as follows:

1. For the decay $\bs$, the fourth generation of quarks may provide more than 
an order of magnitude enhancement in $\acp(q^2)$ in the high-$q^2$ region 
(for $m_{t'}>400\,\rm GeV$), whereas practically no enhancement in the 
low-$q^2$  region is obtained. 
Therefore the high-$q^2$ region is more sensitive to new physics
of this kind.

2. For the decay $\bd$, the fourth generation of quarks may provide 
an enhancement up to $6$ times in $\acp(q^2)$ in the high-$q^2$ region. 
While no enhancement is possible in the low-$q^2$ region for $m_{t'}$
around $400\, \rm GeV$,
at large $m_{t'}$ ($>800\, \rm GeV$) the enhancement in both low and 
high $q^2$ region in the SM4 is about $3$ times the corresponding 
SM3 prediction . 
Since the branching ratio in high-$q^2$ region is small  compared to 
the one in the low-$q^2$ region, the low-$q^2$ region becomes more
attractive at large $m_{t'}$.

3.  For both the decays $B \to (X_s,\,X_d)\, \mu^+ \, \mu^-$, 
the effect of increasing $m_{t'}$ is to increase (decrease)
the values of $A_{\rm CP}(q^2)$ in the low-$q^2$ (high-$q^2$) region.
At large $m_{t'}$, the value of $\acp(q^2)$ is almost independent
of $m_{t'}$.

For a branching ratio of $\sim 10^{-6}$, 
a measurement of a CP asymmetry of $1\%$ at the $3\sigma$ level would require $\sim 10^{10}$ B mesons. Hence
the measurement of a CP asymmetry at the level of a few per cent should be feasible 
at the future colliders like Super-$B$ factories \cite{Browder:2008em,Bona:2007qt}.
Any enhancement observed beyond the standard model, combined with its
$q^2$-dependence, can offer clues about the nature of new physics
involved.

%%%%%%%%%%%%%%%%%%%%%%%%%%%%%
\acknowledgments

A.D. would like to thank C. S. Kim for useful discussions.

\end{document}